\documentclass[a4paper,11pt]{article}
\usepackage{pos}
\usepackage[]{subcaption}
\usepackage{lineno}
\newlength{\bibitemsep}
\newlength{\bibparskip}
\let\oldthebibliography\thebibliography
\renewcommand\thebibliography[1]{%
  \oldthebibliography{#1}%
  \setlength{\parskip}{\bibitemsep}%
  \setlength{\itemsep}{\bibparskip}%
}
\newenvironment{myeq}
    {\setlength{\abovedisplayskip}{3pt}
     \setlength{\belowdisplayskip}{3pt}}
     
\newenvironment{myfig}
    {\setlength{\belowcaptionskip}{0pt}}

\title{Advanced Analysis of Night Sky Background Light for SSTCAM}

\author*[a,b,1]{S.T. Spencer}
\author[c,1]{J.J. Watson}
\author[c,1]{G.Giavitto}
\author[a,1]{G. Cotter}
\author[d,1]{R. White}

\affiliation[a]{Department of Physics, University of Oxford, Denys Wilkinson Building, Keble Road, Oxford, OX1 3RH, U.K.}

\affiliation[b]{ECAP, Friedrich-Alexander Universit{\"a}t Erlangen-N{\"u}rnberg, Nikolaus-Fiebiger-Str. 2, 91058 Erlangen, Germany}

\affiliation[c]{Deutsches Elektronen-Synchrotron, Platanenallee 6, 15738 Zeuthen, Germany}
\affiliation[d]{Max Planck Institut f{\"u}r Kernphysik, Saupfercheckweg 1, 69117 Heidelberg, Germany}

\note{for the CTA SSTCAM Project}
\emailAdd{samuel.spencer@fau.de}

\abstract{Night Sky Background (NSB) is a complex phenomenon, consisting of all light detected by
Imaging Atmospheric Cherenkov Telescopes (IACTs) not attributable to Cherenkov light emission.
Understanding the effect of NSB on cameras for the next-generation Cherenkov Telescope
Array (CTA) is important, as it affects the systematic errors on observations,
the energy threshold, the thermal control of the cameras and the ability of the telescopes to
operate under partial moonlight conditions. This capacity to observe under partial moonlight
conditions is crucial for the CTA transient science programme, as it substantially increases
the potential observing time. Using tools initially developed for H.E.S.S. (in combination with
the prototype CTA analysis package \textit{ctapipe}) we will present predictions for the NSB present
in images taken by the CTA Small Sized Telescope Camera (SSTCAM), showing that
SSTCAM will likely be able to meet the associated CTA requirements. Additionally, we
calculate the potential observing time gain by operating under high NSB conditions.}

\FullConference{%
  7th Heidelberg International Symposium on High-Energy Gamma-Ray Astronomy (Gamma2022)\\
  4-8 July 2022\\
  Barcelona, Spain\\}

\begin{document}
\setlength{\abovecaptionskip}{15pt}

\setlength{\belowcaptionskip}{-12pt}

\maketitle

\section{Introduction}
The upcoming Cherenkov Telescope Array (CTA) $\gamma$-ray observatory will consist of two sites to cover the entire sky; a northern site on La Palma and a southern site on Cerro Paranal in Chile. The southern site will include at least 37 Small Size Telescopes (SSTs) equipped with Silicon Photomultiplier (SiPM) cameras (SSTCAMs), designed to probe the higher end of the IACT energy range above $\mathrm{10\,TeV}$. The ability for these cameras to operate reliably under partial moonlight and other high NSB observing conditions is critical to the transient and multi-messenger/multi-wavelength science goals of CTA, as the potential for astrophysical discoveries scales with the possible observing time. Operating under high NSB conditions significantly increases this. Notably, the detection of prompt TeV emission from the GRB 190114C (the first such detection from the ground) by MAGIC \cite{magicGRB} was under high moonlight conditions, with NSB at approximately 6 times nominal level. However, such observations increase the energy threshold and the potential for systematic error (particularly on flux normalisation) \cite{magicmoon}. In our case this error will likely originate from increased day-to-day NSB rate fluctuation during moonlight observations \cite{magicmoon}. Studying high NSB observations can also inform the design of the temperature control system (as the associated rise in camera trigger rate heats the camera) and the SiPM selection process for SSTCAM.  The majority of previous studies concerning the instrumental response to NSB for SiPM-based Cherenkov cameras have primarily focused on NSB's effect on single SiPM pixels (e.g. \cite{2msipm}); we evaluate both the spatial and temporal properties of NSB across the camera plane for the first time. We also consider the implications of NSB on the operation strategy for SSTCAM's near-real-time LED flasher calibration system (which shines calibrated LED light onto the SiPMs).

\section{Model}
\label{sec:intro:model}

The \textit{sim\_telarray} instrument simulation package \cite{simtel} (used for CTA productions) supports NSB maps as an input parameter. It can also simulate illumination of pixels at an infinite distance from the camera, replicating the effect of stars. But for reasons of computational efficiency, it does not calculate where or how bright those stars might be, nor realistically simulates the effect of moonlight. As a result, we chose to model NSB by expanding on the capabilities of the \textit{nsb} package \cite{nsb}, originally developed by M. Buechele and colleagues in H.E.S.S.. This package assumes that NSB comes from two sources. The first is starlight (using \textit{Gaia} \cite{gaiadr1a} and \textit{Hipparcos} \cite{hipparcos} data), and the second is sky brightness. The latter is semi-analytically approximated (assuming a combination of moonlight and a local extinction coefficient) using a model similar to that in Krisciunas and Schaeffer \cite{Krisciunas}. The \textit{nsb} package makes use of the healpix package to produce skymaps, with which the combined NSB model is expressed using the following quantities. Firstly, $f(\rho)$ is the scattering function, which depends on lunar great circle separation angle $\rho$ as
\begin{myeq}
\begin{equation}
    f(\rho) = 10^{A}\times (1.06+\cos^2\rho)+10^{(B\,-\,\rho/40)}+C \times 10^7 \times \rho^2\ .
    \label{eq:f}
\end{equation}
\end{myeq}
\noindent In this form $f(\rho)$ takes into account both Rayleigh scattering from atmospheric gases and Mie scattering from atmospheric aerosols, with $A$ and $B$ being fitted coefficients. The additional $C$ parameter in Equation \ref{eq:f} was added by the authors of the \textit{nsb} package to account for the relative brightness of the sky to stars. Secondly,
\begin{myeq}
\begin{equation}
    I_{M} (\alpha_{M}) = 10^{-0.4\,\times\,(3.84\,+\,0.026\,\times\,|\alpha_{M}|\,+\,4\,\times\,10^{-9}\,\times\,\alpha_{M}^4)}\ ,
    \label{eq:Im}
\end{equation}
\end{myeq}
\noindent is the illuminance of the Moon in foot-candles ($\mathrm{lm/ft^2}$) as a function of lunar phase angle $\alpha_M$ (subscript $M$ referring to lunar values). Then,
\begin{myeq}
\begin{equation}
    X(Z)=\left.\begin{cases} (1-0.96\sin^2 Z)^{-0.5} & \mbox{, if }  Z\leq\pi/2 \\ (1 - 0.96 \times 1)^{-0.5} & \mbox{, otherwise } \end{cases}\right\}
\end{equation}
\end{myeq}
\noindent is the optical pathlength along the line of sight in units of air masses, where $Z$ is the source zenith angle, and $B_M$ is the surface brightness from moonlight given by
\begin{myeq}
\begin{equation}
    B_{M}(\rho,Z,Z_M,\alpha_M)=f(\rho) \times I_{M}(\alpha_M)\times 10^{-0.4\,k\,X(Z_{M})} \times [1\,-\,10^{-0.4\,k\,X(Z)}]\ ,
\end{equation}
\end{myeq}
\noindent where $k$ is a constant. $B_{Sky}$ is the intrinsic surface brightness of the sky without the moon present, given by 
\begin{myeq}
\begin{equation}
    \begin{split}
    B_{Sky}(Z)=\left. \begin{cases} B_0\times X(Z) \times 10^{-0.4\,k(X(Z)\,-\,1)} & \mbox{, if }  Z\leq\pi/2 \\ 0 & \mbox{, otherwise } \end{cases}\right\}\ ,
    \end{split}
    \label{eq:Bs}
\end{equation}
\end{myeq}
\noindent where $B_0$ is a constant representing the brightness of the sky at the zenith. Finally, $B_{Total}$ is the total brightness for a given pixel on the sky given by
\begin{myeq}
\begin{equation}
    B_{Total}=B_{M}+B_{Sky}+B_{Gaia},
    \label{eq:Bt}
\end{equation}
\end{myeq}
\noindent where $B_{Gaia}$ is the brightness of stars in the healpix pixel. $B_{Total}$ is expressed here in nanoLamberts (nLb), where a brightness in nLb $B$ relates to magnitudes per square arc second in the V band $m_V$ through $B=34.08\exp(20.7233-0.92104\,m_V)\,\mathrm{nLb}$ \cite{Krisciunas}. Some coefficients in the model (such as the exponent in Equation \ref{eq:Im}) are fit to data from Mauna Kea; these will need to be updated once data from the CTA-South site becomes available. 

It should be noted that we neglect potential NSB contributions from zodiacal light, light from population centres, stray reflections from the ground and moonlight reflecting from the secondary mirror, light from satellites and the E-ELT laser guide stars (though all of these should be much smaller than the differences between astronomical dark time and full moonlight). To convert between nLb observed per-pixel (calculated using aperture photometry) and a measured photon rate in Hz/pixel, we use a wavelength-independent model; the SSTCAM front window will heavily suppress emission line NSB that would otherwise mandate a wavelength-dependent analysis. As the results from the \textit{nsb} package cover the \textit{Gaia} Blue Photometer (BP) wavelength range ($\mathrm{330-680\,nm}$) \cite{gaiadr1a}, this implies $B \mathrm{(photons/(ns\ sr\ nm\ m^2) )=}B \mathrm{(nLb)}/(10^4 / \pi \times E \times (680-330))$, where $E=hc/(\mathrm{505\,nm)}$. We then multiply this by the range $\mathrm{300-550\,nm}$ (SSTCAM's wavelength range), an assumed 40\% photon detection efficiency (the peak value for the SiPMs used in the earlier CHEC-S prototype), the solid angle subtended by a pixel ($\mathrm{8 \times 10^{-6}\,sr}$), the mirror area ($\mathrm{7.3\,m^2}$) and the telescope transmission ($0.85$) to get an NSB rate in Hz. The current we ultimately measure in the camera front-end electronics as a result of this photon flux is dependent upon the gain of the SiPMs, necessitating the use of the flasher system to determine how the gain has changed.

\section{Results}
\begin{figure}[t!]
\begin{minipage}{0.9\linewidth}\centering
\begin{myfig}
\subcaptionbox{Dark empty field.}{\includegraphics[width=0.45\linewidth]{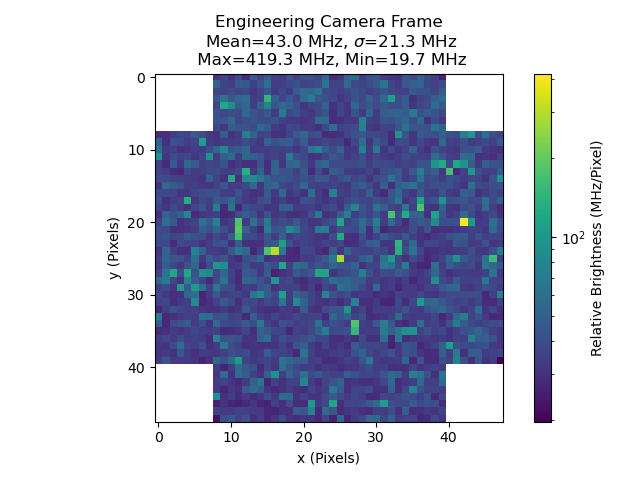}}
\subcaptionbox{No moonlight Eta Carinae.}{\includegraphics[width=0.45\linewidth]{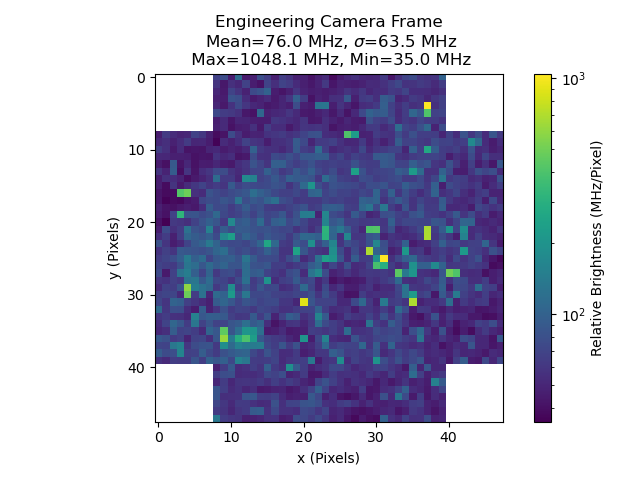}}
\subcaptionbox{Half moonlight Eta Carinae.}{\includegraphics[width=0.45\linewidth]{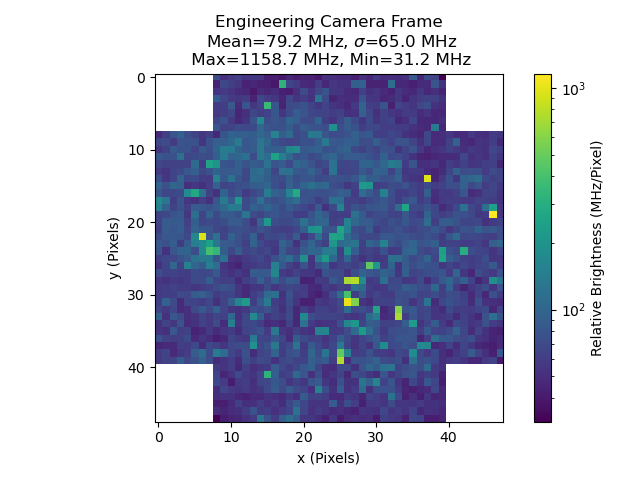}}
\subcaptionbox{Full moonlight Eta Carinae.}{\includegraphics[width=0.45\linewidth]{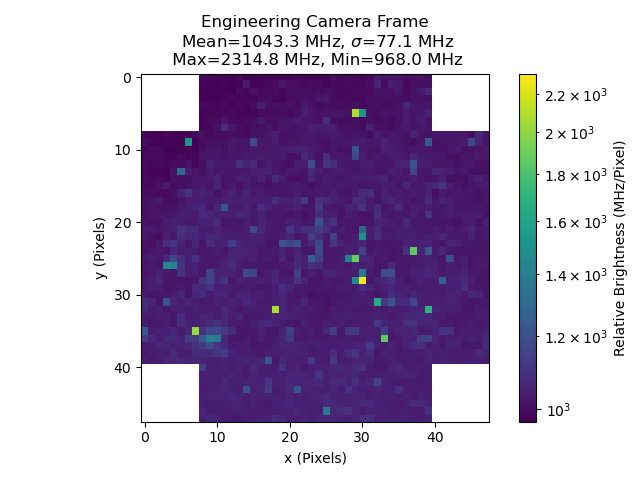}}
\end{myfig}
\caption{Per pixel NSB values for Eta Carinae under the four scenarios described. To extract pixel values from the \textit{nsb} package skymaps we used the \textit{ctapipe} package \cite{ctapipe-icrc-2021}. Please note the differing colourbar scales.}
\label{fig:etacarpixel}
\end{minipage}
\end{figure}

\begin{figure}[t!]
\begin{minipage}{\linewidth}\centering
\subcaptionbox{}{\includegraphics[width=0.46\linewidth]{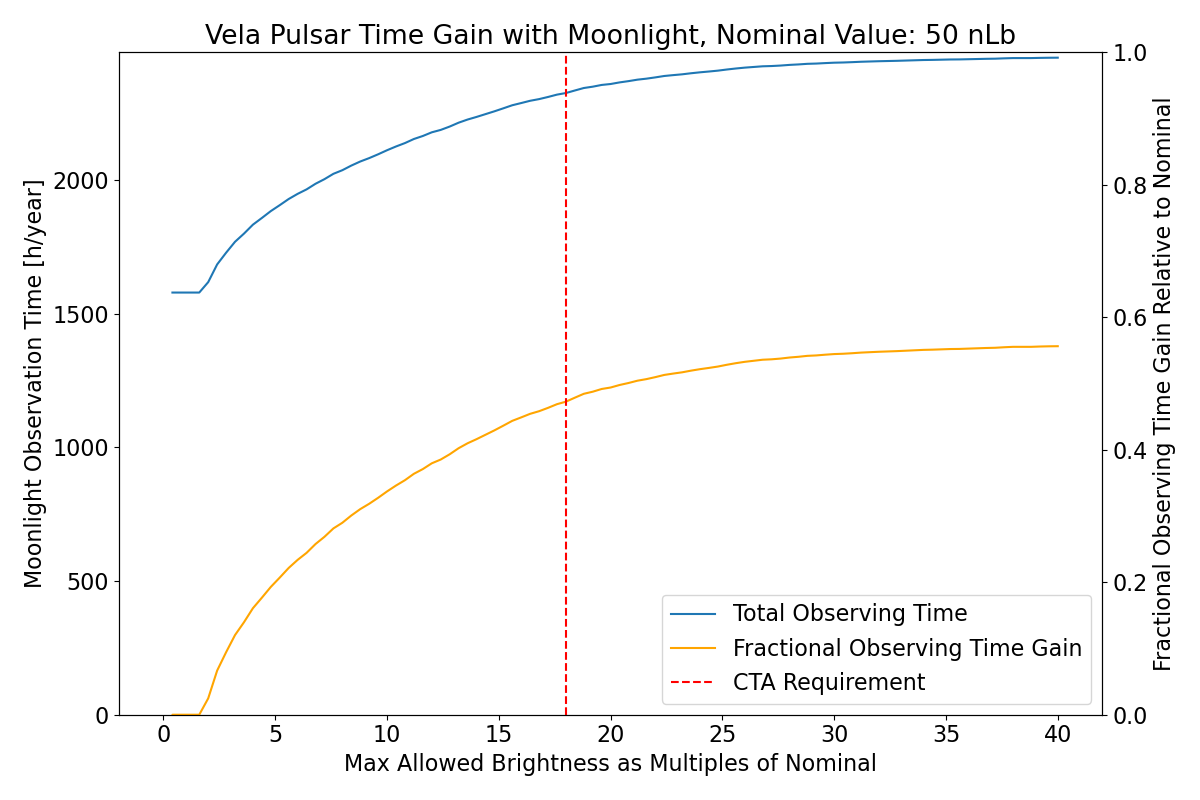}}
\subcaptionbox{}{\includegraphics[width=0.46\linewidth]{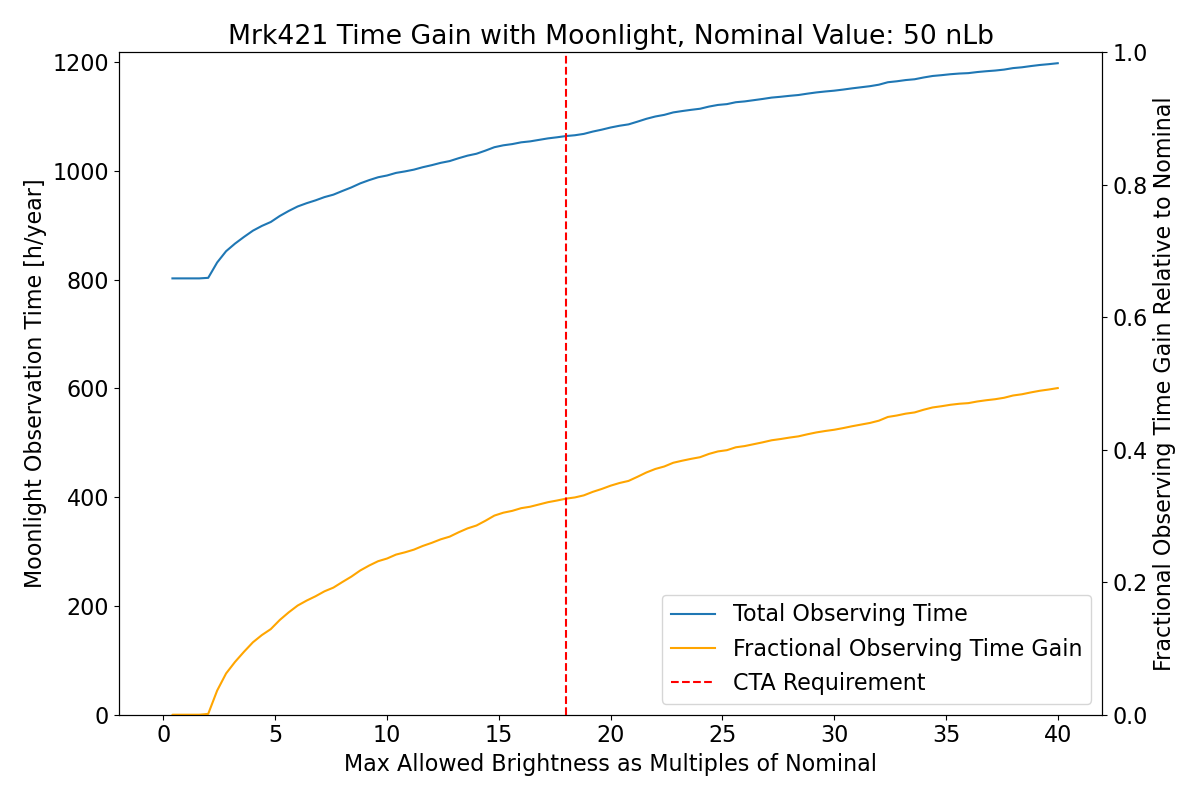}}
\caption{Observing time gains possible for a galactic and extragalactic source as a function of nominal NSB.}
\label{fig:obstimegainvelapulsar}
\end{minipage}
\end{figure}

As a reasonable worst-case scenario for the expected rates of NSB per SiPM pixel, we consider four observing scenarios of the colliding wind binary Eta Carinae. This is considered to be a particularly difficult source to observe with IACTs given the high stellar density in the region. To begin, we consider a `dark' field, at the same altitude as Eta Carinae but differing azimuth, for comparison. Secondly we consider observing Eta Carinae during astronomical dark time (at the same observing time as the dark field). Thirdly, we observe the Eta Carinae region with half moonlight present, when the moon is above the horizon and has 0.53 Fractional Lunar Illumination (FLI). Finally, we observe the region under full moonlight. The full moon scenario represents the worst possible observing conditions for Eta Carinae (and by extension the worst observing conditions possible), with the moon both at 1.0 FLI and well above the horizon. The results for these scenarios are shown in Figure \ref{fig:etacarpixel}.
The \textit{nsb} package also provides a function to make plots of the total possible observation time for a point source as a function of NSB threshold.  To consider the NSB across the entire camera plane, we normalise these brightness values to the mean NSB value in our dark frame.  To consider potential differences between galactic and extragalactic sources, we performed this calculation for the Vela pulsar and the blazar Markarian 421 (Figure \ref{fig:obstimegainvelapulsar}) over a year. Operating to the CTA requirement shows an approximate 30\% observing time gain over operating purely at nominal NSB values; the gain is slightly more pronounced for galactic sources given the higher stellar density in the galactic plane.

\begin{figure}[ht]
\begin{minipage}{\linewidth}\centering
\subcaptionbox{40s flasher calibration intervals.}{\includegraphics[width=0.46\linewidth]{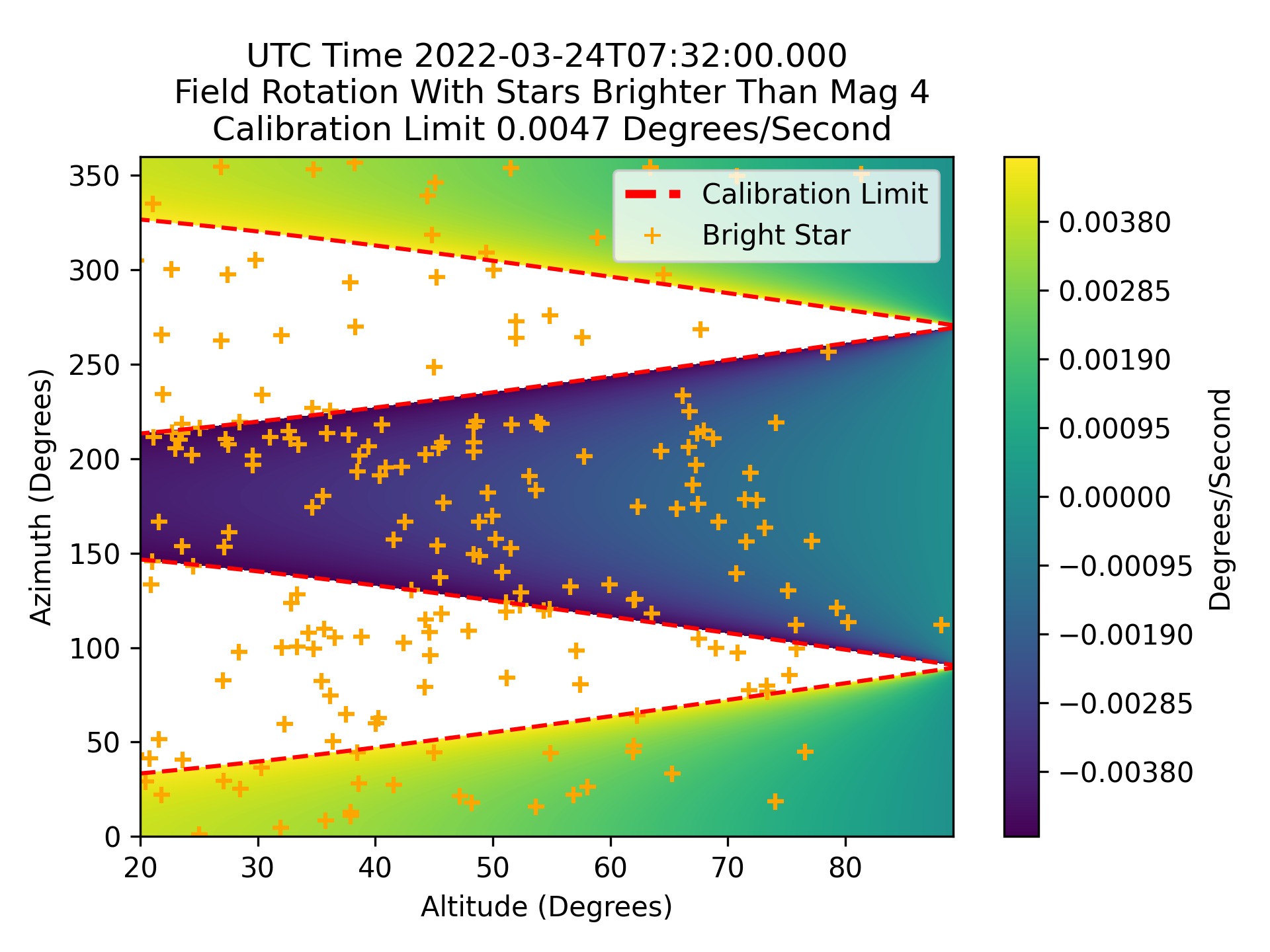}}
\subcaptionbox{10s flasher calibration intervals.}{\includegraphics[width=0.46\columnwidth]{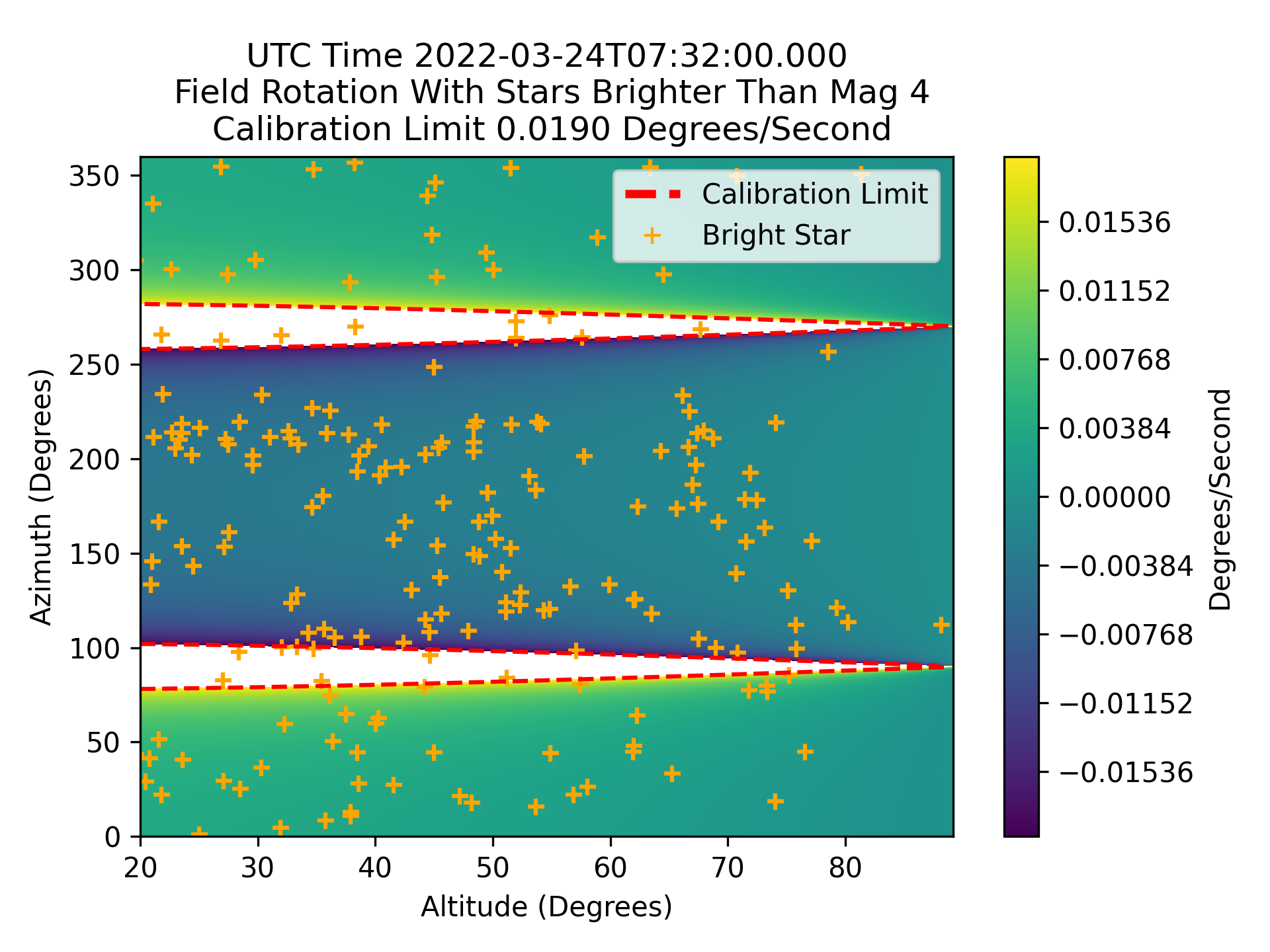}}
\caption{Field rotation with flasher calibration limits for given calibration intervals.}
\label{fig:rot40s}
\end{minipage}
\end{figure}
Bright stars are required to be detected by the SiPMs for the purpose of pointing calibration in the SSTCAM slow-signal chain, but this could potentially affect event reconstruction in the fast-signal chain. This is a problem unique to SSTCAM, and requires that the effect of bright stars upon the telescope calibration is well understood. Given that the angular field rotation of such a bright star $R$ can be described as a function of Latitude ($Lat$), Altitude ($Alt$) and Azimuth ($Az$) using $R(Lat,Alt,Az)=\mathrm{15.04}\cos(Lat)\cos(Az)/\cos(Alt)\,\,\mathrm{^{\circ}/hour}$, it's possible to extract the number of required flasher calibration pulses to reach a mean flasher error level. If we assume a mean illumination level from the flasher of $50\,\mathrm{photoelectrons}$, and that the excess noise factor (a property of the SiPMs used that describes the physical limit for the best possible SiPM charge resolution) is 1.4, then we need $40\,\mathrm{s}$ (i.e. $400\,\mathrm{pulses}$) to reach a 1\% error on the mean flasher level per pixel. This is the needed flasher calibration level for SSTCAM to be able to meet the intensity resolution requirements of CTA for high-amplitude signals. The limits on stellar rotation rate occur at the point at which stars cause a gain change in a SiPM pixel that happens more rapidly than can be calibrated for by injecting flashes at a fixed rate. 

Figure \ref{fig:rot40s} shows the areas on the sky where a 1\% flasher error can be achieved given this model of angular field rotation, along with the positions of stars brighter than fourth magnitude from \textit{Hipparcos} \cite{hipparcos}. This stellar rotation causes a slight error in calibration along the North-South axis, but this is tolerable given the rate at which stars move across the sky. By doubling the error budget and reducing the intervals between the flashing procedure, one can achieve much greater calibrated sky coverage, with only a handful of bright stars falling outside the calibrated range. However, these results suggest that the current flasher calibration plan for SSTCAM of using the flasher to generate 100-200 photons with nanosecond precision will be acceptable.

\section{Conclusions}
The results from our investigation demonstrate that SSTCAM operations will be minimally affected by bright moonlight conditions. For the dark field we compute a photon rate of $\mathrm{43\,MHz}$, which agrees (within Poisson error) with a calculation performed by K. Bernl{\"o}hr who obtained $\mathrm{41.9\,MHz}$ (based on scaling the Benn and Ellison spectrum) \cite{BandE,konradpc}. Even in our worst observing scenario, the average photon rate is within SSTCAM observing requirements. However, the combination of moonlight and a bright stars means pixels with an NSB rate of more than a few GHz will still need to be disabled and removed from Cherenkov analysis (to manage SiPM heating). In the dimmer case of illumination at 
a rate of around 1\,GHz, it is likely such pixels will only need to be removed from the camera trigger. As such, the potential for SSTCAM science during high NSB conditions remains promising. 

\section{Acknowledgements}
This work was conducted in the context of the CTA Analysis and Simulations Working Group, and this paper has been through review by the CTA Consortium. We gratefully acknowledge financial support from the agencies and organisations listed here: \url{http://www.cta-observatory.org/consortium_acknowledgments}. STS has been supported by an STFC PhD scholarship, the University of Oxford, and Deutsche Forschungsgemeinschaft (DFG, German Research Foundation) Project Number 452934793.

\bibliographystyle{JHEP}
\bibliography{references}
\end{document}